\documentclass[12pt]{article} 

\usepackage{graphicx}
\usepackage{hyperref,url}
\usepackage{color}
\usepackage{amssymb,latexsym}
\usepackage{amsmath,amsthm}
\usepackage{enumerate}
\usepackage{tikz}
\usetikzlibrary{arrows.meta}
\usetikzlibrary{shapes,positioning,bending}
\usepackage{amsfonts}
\usepackage{dsfont}
\usepackage{caption}
\usepackage{subcaption}
\usepackage[makeroom]{cancel}

\usepackage{adjustbox} 
\usepackage{soul}
\usepackage[final]{microtype}

\usepackage{authblk} 
\usepackage{xcolor}
\usepackage{comment}
\usepackage[title]{appendix}

\usepackage[absolute]{textpos} 

\usepackage{geometry} 
\usepackage[font=small,labelfont=bf]{caption} 
\usepackage{palatino}

\usetikzlibrary{arrows}
\usetikzlibrary{chains,matrix,scopes}

\usepackage[utf8]{inputenc}
\usepackage[english]{babel}

\theoremstyle{plain}

\theoremstyle{definition}

\newtheorem{definition-theorem}[theo]{Definition-Theorem}

\theoremstyle{remark}

\newcommand{\bit}{\textrm{bit}} 
\newcommand{\bits}{\textrm{bits}} 
\newcommand{\NAT}{\textrm{nat}} 

\newcommand{\Ybit}{\textrm{Ybit}}
\newcommand{\J}{{\textrm J}}
\newcommand{\K}{{\textrm K}}

\makeatletter
\renewcommand{\p@subfigure}{\thefigure.}
\makeatother

\def \be {\begin{equation}}
\def \ee {\end{equation}}
\def \bes {\begin{equation*}}
\def \ees {\end{equation*}}
\def \baa {\begin{align}}
\def \eaa {\end{align}}
\def \baas {\begin{align*}}
\def \eaas {\end{align*}}
\def \bea {\begin{eqnarray}}
\def \eea {\end{eqnarray}}
\def \beas {\begin{eqnarray*}}

\title{Temperature as Joules per Bit\thanks{To appear in the American Journal of Physics, \url{https://doi.org/10.1119/5.0198820} ~ ~ ~  ~ ~ ~  ~ ~ ~  
\makebox[207pt]{$^\dagger$Authors are listed in alphabetical order.}}}

\author[1]{Charles Alexandre Bédard}
\author[2]{Sophie Berthelette}
\author[3,\,4,\,5]{\\Xavier Coiteux-Roy}
\author[1,\,6]{Stefan Wolf\hspace{19pt}\textsuperscript{$\dagger$}\hspace{-23pt}}

\affil[1]{\small{Faculty of Informatics, Universit\`a della Svizzera italiana, Lugano, Switzerland}}
\affil[2]{\small{Département de physique, Université de Montréal, Montréal, Canada}}
\affil[3]{TUM School of Natural Sciences, Technical University of Munich, Garching, Germany}
\affil[4]{TUM School of Computation, Information and Technology, Technical University of Munich, Garching, Germany}
\affil[5]{Munich Center for Quantum Science and Technology (MCQST), Munich, Germany}
\affil[6]{\small{Facolt\`a indipendente di Gandria, Gandria, Switzerland}}

\date{February 2025}

\begin{document}
\maketitle
\thispagestyle{empty}

\begin{abstract}
\noindent In statistical mechanics, entropy is defined as a fundamental quantity. 
However, its unit, $\J/\K$, involves that of temperature, which is only subsequently defined~--- and defined in terms of entropy.
This circularity arises with the introduction of Boltzmann's constant into the very expression of entropy.
The $\J/\K$ carried by the constant prevents entropy from finding a unit of its own while simultaneously obfuscating its informational nature.
Following the precepts of information theory, we argue that entropy is well measured in bits 
and coincides with information capacity at thermodynamic equilibrium.
Consequently, not only is the temperature of a system in equilibrium expressed in $\J/\bit$, but it acquires a clear meaning: 
It is the cost in energy to increase its information capacity by 1~bit.
Viewing temperature as joules per bit uncovers the strong duality exhibited by Gibbs long ago between available capacity and free energy.
It also simplifies Landauer's cost and clarifies that it is a cost of \textit{displacement}, not of erasure. Replacing the kelvin with the bit as an SI unit would remove Boltzmann's constant from the seven defining constants.
\end{abstract}

\maketitle
\thispagestyle{empty}

\newpage
\section{Motivation}\label{sec:mot}

Entropy is a fundamental quantity in statistical mechanics, and its SI unit is~$\J/\K$. 
Yet, 
\emph{in statistical mechanical terms}, what does a~$\J/\K$ represent?
Tentative answers must confront the circularity mentioned in the abstract, where entropy is defined first yet incorporates a logically posterior temperature unit.
Articulating 
entropy in terms of an as-yet-undefined unit demands an explanation that necessarily breaks with entropy's otherwise simple definition.
It must foreshadow a logically posterior definition, like \mbox{$T = \left( \partial S / \partial E \right)^{-1}$}, or theorem, like equipartition.
Moreover, to think in joules per kelvin introduces into our view of entropy all the parochial dependencies encompassed in the development of the Kelvin scale. 
For instance, this includes the aggregate behavior of water and the number 100 --- and with it, our physiology: the number of our fingers.

The flipside of the aforementioned circularity is that, upon defining temperature from energy and entropy, the kelvin conspicuously crops up alone, as if it were the unit of a fundamental quantity.
For instance, Schroeder~\cite{schroeder2000introduction} wrote: ``Thanks to the factor of Boltzmann's constant in the definition of entropy, the slope $\partial S / \partial E$ of a system's entropy vs. energy graph has the units of $(\J/\K)/\J=1/\K$. If we take the reciprocal of this slope, we get something with units of kelvins, just what we want for temperature.''

In Section~\ref{sec:his}, we first explain how the circular inconsistency regarding the entropy units arises in statistical mechanics as a result of its historical ties to phenomenological thermodynamics.
The significant role of information in thermodynamics is emphasized in Section~\ref{sec:infothermo}, which motivates the information-theoretic account of entropy.
As explained in Section~\ref{sec:eic}, at thermodynamic equilibrium, entropy amounts to information capacity.
In Section~\ref{sec:temperature}, we show how temperature arises as a bridge between energy and information capacity, and how it acquires a clear meaning when formulated in units of joules per bit.
Not only does it uncover the notion of available information capacity completely analogous to free energy, but it also sheds light on Landauer's erasure. We also discuss how our proposal alters the current SI.

\section{The Ties of History}\label{sec:his}
\emph{``Careful, that's hot!''}
Temperature is one of the most intuitive physical concepts, as we can vividly feel it through our skin.
Arguably, \emph{temperature} is more intuitive than \emph{energy}; and definitely more than \emph{entropy}.
Perhaps unsurprisingly, the historical development of the three concepts has also followed that order.

The first thermometers were built in the 17\textsuperscript{th} century, by the end of which some precursor of kinetic energy, the so-called \emph{vis viva}, was introduced by Leibniz~\cite{leibniz1695specimen}.
As Young~\cite{young1807course} coined the term energy at the beginning of the 19\textsuperscript{th} century, its existence in various forms was quickly realized.
Notably, in 1824, Carnot~\cite{carnot1824reflexions} studied engines that could use energy of the \emph{caloric} type to produce energy of the \emph{work} type.
Clausius~\cite{clausius1850ueber} then characterized the form of energy that is inevitably lost during a thermodynamical cycle in such engines.
Only then did he introduce entropy, a monotonically increasing quantity which assured the irreversibility of some processes.

The formalization of these concepts at a macroscopic level, rooted in empirical observations, established the first theory of heat: \emph{thermodynamics}.\footnote{While in this work we use the term ``thermodynamics'' in its general sense encompassing also statistical mechanics, the present section assumes ``thermodynamics'' to specifically denote the phenomenological and macroscopic framework that remains agnostic to any underlying microscopic physics, i.e.\ not statistical mechanics.} 

The first theory of heat is a theory of thermal processes relating macroscopic quantities. Its starting point is temperature.
Indeed, the zeroth law entails grouping systems into equivalence classes based on thermal equilibrium.
Temperature then serves as a real-valued label for these equivalence classes, endowing them with an ordering that dictates the direction of possible heat flow.
Clausius' formula for entropy is~$\Delta S={Q}/{T}$, where~$Q$ is the amount of heat entering or escaping the system and~$T$ is the absolute temperature of the system.

The second theory of heat, \emph{statistical mechanics}, departs from the first by taking into consideration the underlying atomic structure. Statistical mechanics depends on microstates, and it takes their count $\Omega$, the number of possible configurations, as more fundamental than temperature.
Following this precept, Boltzmann's entropy could be, as it was in Boltzmann's writings~\cite{anirban2021constant}, simply without constant: 
$$
\xcancel{k_B}\,  \ln\Omega \,.
$$
Boltzmann's entropy in this raw form is a combinatorial object, a log-count, a very different quantity than the entropy of Clausius.
Had~$\ln \Omega$ been given a life of its own, perhaps Boltzmann's entropy had eventually found its own unit --- in hindsight the natural unit of information (nat).

But it did not. 
Planck prefactored $\ln \Omega$ with what is now known as Boltzmann's constant, $k_B$.
The constant harmonizes Boltzmann's statistical mechanical entropy with Clausius's thermodynamic entropy, both in $\J/\K$, while also assuring a statistical mechanical temperature 
in kelvins.

We suggest that this introduction of the constant $k_B$ \emph{into} Boltzmann's entropy, as important as it was for the early development of statistical mechanics, tied the infant theory of statistical mechanics to its predecessor, thermodynamics. 
However, the theory of thermodynamics should be obtained as a limiting case of statistical mechanics, and not be used as an overarching framework for statistical mechanics.
This is because statistical mechanics supersedes thermodynamics; it claims more explanatory power.
For instance, it yields fluctuation results, it accommodates quantum statistics,
it explains phase transitions from first principles, and it better predicts low temperature behaviors.
The chronological discovery of ideas does not, therefore, amount to their logical priority.

\section{Information in Thermodynamics}\label{sec:infothermo}

Crucial insights and results followed the introduction of information-processing considerations into the picture of thermodynamics.
For modern reviews, see Refs~\cite{maruyama2009colloquium, parrondo2015thermodynamics}.
A serious challenge to the second law was proposed by Maxwell~\cite{maxwell1871theory, leff1990maxwell, leff2002maxwell} with his famous demon who sorts a system via measurements and manipulations, thereby reducing its entropy.
Szilard's engine~\cite{szilard1929entropieverminderung} has become the canonical system on which a demon is imagined to perform its puzzling manipulations.
It consists of a one-molecule gas container that can be bisected by a partition, which thereupon acts as a bidirectional piston.
After inserting the partition, the demon observes on which side the molecule is trapped, and then expands adiabatically the piston in the opposite direction.
This extracts~$k_B T \ln 2$ of free energy from the molecule. 
After thermalization with the environment, the process can be repeated, \emph{if} the demon has kept his ability to measure and act on the system. 
Szilard suggested that the second law should be saved by the act of measurement of the demon, which, he thought, should unavoidably create $k_B\/ T \ln 2$ of heat, namely, just enough to compensate.
The idea to tie a thermodynamic cost to the act of measurement per se has been followed by many \cite{von1955mathematical, gabor1961iv, brillouin1962science}, 
and was likely stimulated by the confusion that arose with measurement in quantum theory.
 
In 1961, Landauer~\cite{landauer1961irreversibility} correctly identified information erasure --- and not measurement --- as the precise thermodynamically irreversible step which needs to be compensated by heat dissipation.
The erasure of one bit of information must be accompanied by an amount of~$k_B\/ T \ln 2$ of free energy lost to the environment, or to the non-information-bearing degrees of freedom of a computer. 
The reason for heat dissipation is purely physical: Information cannot be processed independently of real devices, or as Landauer would have it, ``\emph{Information is physical}.''

The fact that erasure has a thermodynamic cost does not \emph{a priori} preclude measurement from also having such a cost.
This concern was part of the larger misconception which held that computing devices should unavoidably involve logical irreversibility, and, with it, heat dissipation.
Upon demonstrating that universal computation can be done via logically reversible steps, Bennett~\cite{bennett1973logical} paved the way for developing thermodynamically reversible models of computation, the most spectacular of which is perhaps Fredkin and Toffoli's ballistic computer~\cite{fredkin1982conservative}. 
In this light, the status of measurement, a very special kind of computation, has been clarified~\cite{bennett1982thermodynamics}: An apparatus initialized in a ready state can measure non-dissipatively.
This yields a detailed and satisfactory resolution of Maxwell's problem: With an initialized memory, the demon \emph{can} measure the system and act on it to reduce its entropy.
But this is no paradox, as it merely displaces the entropy of the system onto its memory.
To operate in a cycle, the demon needs to reset the memory to its initial state, that is, to get rid of the information stored --- an erasure which, by Landauer's bound, dissipates a quantity of heat greater or equal to the entropy reduction of the system times the environmental temperature.

The logical reversibility of dynamical laws applies to all physical systems, including those with the ability to store information and act based on it. 
Information-processing agents cannot avoid the second law.
Yet, the point of view of information processing has offered
limits to what can or cannot be done, thermodynamically.
In particular, Landauer’s erasure cost can be considered one of the many expressions of the second law, which, as eloquently stated by Schumacher~\cite{schumacher2011quantuminfo}, mandates that ``No physical process has as its sole result the erasure of information.''

The resolution of Maxwell's demon has been highly influential in promoting the role of information theory in thermodynamics. 
The scientific literature on the topic has been booming and ramifying in many ways.
The advent of Shannon's mathematical theory of communication~\cite{shannon} generated insights~\cite{jaynes1957information, jaynes1957information2, hanel2014multiplicity} (and debates~\cite{shimony1985status, toffoli2016entropy}) on the nature of entropy. 
Algorithmic information theory~\cite{solomonoff1964formal, Kolmogorov1965, chaitin1966length} permits a quantification of information based on individual objects, yielding more sophisticated notions of entropy~\cite{zurek1989algorithmic, gell2010effective} and macrostates~\cite{CCC}.
The significance of quantum information was realized~\cite{holevo1973some, brassard1984quantum, deutsch1985quantum,
vedral2006introduction} and incorporated into various thermodynamical analyses~\cite{lubkin1987keeping, plenio2001physics, zurek2003quantum, rio2011thermodynamic, 
faist2015minimal, 
goold2016role}.
And axiomatic reconstructions of thermodynamics have been suggested~\cite{lieb1999physics, marletto2016constructor, hulse2018axiomatic}.

\section{Thermodynamic Equilibrium: Entropy as Information Capacity} \label{sec:eic}
Energy is a fundamental physical quantity whose conservation principle has permitted significant theoretical advancements and has been proven to be repeatedly consistent with experimental tests.
In a conversation about thermodynamics, we shall be satisfied with its SI unit, the joule ($\J$), itself expressed in terms of the kilogram, the meter, and the second.

However, we expressed the problematic circularity
with the SI unit of entropy as measured in $\J/\K$ (Section~\ref{sec:mot}), and advocated instead for the information-theoretic view of entropy (Section~\ref{sec:infothermo}).
Many such entropy measures have been proposed, most of which are tied to bits.  
For instance, Shannon entropy~$H(X)$ is the expected number of bits required to communicate the outcome of a random variable~$X$ in an optimal prefix code~\cite{shannon}.
More convoluted measures were proposed; yet, for our purposes, we step back on what is perhaps the simplest: information capacity.

\medskip

The \emph{information capacity}~$S(E)$ (or simply the \emph{capacity}) of a system at a given energy~$E$ is quantified by
$$
S(E) = \log_2 \Omega(E) ~~\bits \,,
$$
where $\Omega(E)$ is the number of possible configurations a system at energy $E$ can adopt,\footnote{
A classic interpretation of $\Omega$ is the cardinality of a macrostate, which here is solely determined by the system's energy.} and a \bit~is the information capacity of a 2-level system.
The logarithm of a microstate count that corresponds to a given constraint is the gist of Boltzmann's entropy. 
By using the base~$2$ instead of the natural logarithm,\footnote{A valid defense of the natural logarithm is its convenience when the machinery of differential calculus is used to elaborate the theory. While true, this is easily resolved by the conversion~$\log_2 \Omega(E) ~~\bits = \ln \Omega(E) ~~\textrm{nat}$.} and prefactoring it by~$1~\bit$ instead of~$k_B$, we obtain a measure of information capacity in bits, corresponding to a given energy~$E$.
Information capacity is a counterfactual property of the system, namely, an assessment of all the configurations in which the system \emph{could be}, and it does not depend on the system's actual configuration.
Nor does it depend on any probabilistic assessment, thereby avoiding the question of subjectivity and agent dependency.

Like a distance in space measured by the number of meters that can fit in that space, the information capacity of a system is measured by the number of bits that can fit in that system. 
In this light,
$S(E)$ is the number of bits required to label all distinguishable states at energy~$E$, or equivalently, the number of bits that can be encoded in the system if it is seen as a storage resource.
At the macroscopic scale, physical systems have a very large number of microstates, which translates to a large value for information capacity. 
The divide between a two-level system and a macroscopic system can be bridged, for instance,
by measuring information capacity in yottabit ($\Ybit$), which corresponds to $10^{24}~\bits$.
This unit assesses more conveniently the information capacity of macroscopic systems.

In modern days, it has become second nature to quantify the storage capacity of devices in $\bits$. 
In a similar way, any physical system can be viewed as an information-storage system.
Compared to the memory of a computing device that has been engineered to be stable in some relevant set of environments, a generic physical system has microstates which, for all practical purposes, cannot be prepared nor maintained in any chosen configuration.
Despite this technological (but not fundamental) impracticality for user interplay, it remains that information can be encoded in physical systems, a quantity which is bounded by its information capacity.
The assumption of thermodynamic equilibrium amounts to viewing a system's entropy as maximal and, therefore, reaching its full capacity.\footnote{The fact that a system at maximal entropy saturates its capacity can be easily recognized when quantifying entropy in the probabilistic setting: In this case, the Gibbs--Shannon entropy $S(\rho)= - \sum_{i=1}^{\Omega(E)}p_i\log_2 p_i$ 
of a distribution $\rho$ over a configuration space of~$\Omega(E)$ possible values is maximal for the uniform distribution, and is then equal to the capacity.}

\section{Temperature as Joules per Bit}\label{sec:temperature}

In this section, we develop the logical implication of taking entropy as fundamental and assigning it its own unit: temperature should have the units of~$\J/\bit$.
The information capacity~$S$ is a function of its internal energy~$E$.
The temperature is obtained as the reciprocal of the slope between information capacity and energy, $T = \left ( \partial S / \partial E\right )^{-1}$, or, when the slope is well-defined, directly as~$\partial E / \partial S$.
Therefore, with entropy measured in \bits, temperature is in $\J/\bit$.

Non-exotic systems are of positive temperature and positive heat capacity, yielding both positive first and positive second derivatives of~$E$ with respect to~$S$, as displayed in~Figure~\ref{traditional}.\footnote{In most contexts, $S(E)$ can be inverted into a function~$E(S)$. We opted for the graph of the latter, so as to have the slope directly equal to the temperature (instead of the inverse temperature).}
In thermodynamic contexts, where large systems are concerned, $1~\bit$ is negligible compared to the capacity of the system, so the slope~$\partial E / \partial S$ is well approximated by a finite difference. Thus temperature can be interpreted as the increase in internal energy (in~$\J$) required to increase the information capacity by $1~\bit$.

\begin{figure}[ht]
\centering
\begin{subfigure}[t]{0.4 \textwidth}\centering
\begin{adjustbox}{width=\textwidth}
\begin{tikzpicture}
    \draw [thick, <->] (0,6) node [left, rotate=90, yshift=3mm] {Energy ($E$)} -- (0,0) -- (6,0) node [below left] {Information capacity ($S$)};
    
    \draw [very thick] (0.2,0.3) to [out=10, in=-92] (5,6);

    \draw [red, thick] (2.92,1.5) -- (3.4,1.5) -- (3.4,1.94) -- cycle;

    \coordinate (PointOnCurve) at (3.123,1.721);
    \fill (PointOnCurve) circle (2pt);

    \draw (3.15,1.5) node [below] {$\Delta S$};
    \draw (3.4,1.7) node [right] {$\Delta E$};

 		\end{tikzpicture}
		\end{adjustbox}
		\captionsetup{width=0.85\linewidth}  
		\subcaption{Systems usually have a\\
		 \hspace{12pt} concave up shape.}
		\label{traditional}
		\end{subfigure}
		\hspace{0.9cm}
		\begin{subfigure}[t]{0.4 \textwidth}\centering
		\begin{adjustbox}{width=\textwidth}
 		\begin{tikzpicture}
 		\draw [thick, <->] (0,6) node [left, rotate=90, yshift=3mm] {Energy ($E$)}--(0,0)--(6,0) node [below left] {Information capacity ($S$)};
		\draw [very thick] (0.2,0.3) to (6,5);
 		\end{tikzpicture}
		\end{adjustbox}
		\captionsetup{width=0.85\linewidth}  
		\subcaption{A heat bath is a linear\\
		 \hspace{13pt}  idealization.}
		\label{heat}
 	    \end{subfigure}
 			\captionsetup{width=0.9\linewidth}  
 	\caption{Graphs of energy vs\ capacity.}
 	\label{graphs}
\end{figure}

The statistical mechanical definition of temperature has many advantages, one of which is the possibility to make sense of negative temperature~\cite{baumeler2022thermodynamics}.
In joules per bit, it is to be interpreted as the amount of energy that needs to be extracted from the system in order to increase its capacity by one bit.
Furthermore, temperature as such is independent of an \emph{a priori} notion of \emph{heat baths}, which instead can be understood in terms of systems whose $E(S)$ function is of constant slope, namely, of constant temperature, as displayed in Figure~\ref{heat}. 
Such systems can be thought of as idealizations of very large systems or, as close-ups of some $E(S)$ function, which then appears to be linear.
The energy cost to increase the information capacity of a heat bath at temperature~$T$ (in~$\J/\bit$) by~$X$ (in \bits) is~$XT$ (in~\J).
%
However, for systems that are not well approximated by a heat bath, the energy cost for an additional~$x ~\bits$ need not be linear: The energy cost must be integrated over the interval of increased information capacity.

\subsection{Available Energy and Capacity}

In 1873, Gibbs~\cite{gibbs1957method} found out that systems whose entropy is not maximal have \emph{available energy} --- now known as \emph{free energy} --- as the system can be used to produce work.
This energy is extracted from the system by transforming it into a state of lower energy, while conserving its entropy, until the capacity curve is reached.   
It is thus the amount of energy that can be extracted from the system with no need to store an excess of entropy.
In the same breath, Gibbs presents what he calls the \emph{capacity for entropy}, or as we like to view it, \emph{available capacity}.\footnote{Available capacity is also known by the names \emph{negative entropy} (Schrödinger~\cite{schrodinger1944life}) and \emph{negentropy} (Brillouin~\cite{brillouin1953negentropy}).}
It is the number of bits that can be encoded in the system at no extra energy cost.
Available capacity quantifies the amount of structure in the system, like blanks on a tape, or the empty registers in the memory of Maxwell's demon. 
The view advocated here, in which energy and entropy are more fundamental than temperature, highlights well the duality captured by Gibbs between available energy and available capacity. 
Figure~\ref{fig:available} illustrates these quantities in the energy vs capacity plane of an isolated system.

\begin{figure}[h]
    \centering
    \begin{tikzpicture}[scale=0.9]
    \draw [thick, <->] (0,6) node [left, rotate=90, yshift=3mm] {Energy ($E$)} -- (0,0) -- (6,0) node [below left] {Information capacity ($S$)};
    
    \draw [very thick] (0.2,0.3) to [out=10, in=-92] (5,6);

    \coordinate (PointOnCurve) at (3,3.5);
    \coordinate (VerticalProjection) at (3,1.62);
    \coordinate (HorizontalProjection) at (4.4,3.5);

    \draw [dashed, red] (PointOnCurve) -- (VerticalProjection);
    \draw (VerticalProjection) node [above left, align=center, font=\scriptsize, yshift=5.5mm] {Available\\energy};

    \draw [dashed, blue] (PointOnCurve) -- (HorizontalProjection);
    \draw (HorizontalProjection) node [
    align=center, font=\scriptsize, 
    yshift=4mm,
    xshift=-6mm] {Available\\capacity};

    \fill (PointOnCurve) circle (2pt);

    \end{tikzpicture}
    \captionsetup{width=0.9\linewidth}  
    \caption{The point's coordinates are given by the system's entropy and energy. It represents a nonequilibrium state, as its entropy is not maximal, i.e., the point is not on the capacity curve. As an example, it could correspond to the state of an isolated gas in a box,
    where all the particles occupy a smaller region than that of the whole box.
    The vertical red dashed line represents the available energy while the horizontal blue dashed line represents the available capacity.}
    \label{fig:available}
\end{figure}

Moreover, that duality is also captured by idealized systems that are efficient to store either energy or entropy. 
A \emph{battery} is a system designed to keep the energy available; namely, the internal energy of the system can be changed with no significant changes in entropy.
As an example, consider a weight in a gravitational field. 
The (potential) energy of the weight can be easily changed without affecting the information that is encoded in it. 
When changing the position of the weight,
except for the energy, all the intrinsic properties of the system 
are unchanged.
On the other hand, some systems are efficient for storing information (or entropy) with no energy change --- we call them \emph{tapes}. 
Degenerate ground states and ideal hard disk drives are instances of those tapes.

\subsection{Landauer's ``Displacement''}
Be it classical or quantum, information processing follows the logical reversibility of the physical laws of motion.
As a consequence, the information (i.e. entropy) of an isolated system does not decrease spontaneously. 
When considering the entire universe as one isolated system, the conclusion is straightforward: ``No information is ever lost.'' 
This statement has been recognized as a formulation of the second law of thermodynamics~\cite{wolf2018second,baumeler2019free}.

The term ``information erasure'' is, therefore, a misnomer: Information is never \emph{erased} as if free energy had the power of fundamental erasure. 
Rather, information is \emph{displaced}. 
When information leaves the relevant degrees of freedom of an information-storage device, it moves to nearby systems, its environment.
The environment, therefore, should be seen as a memory which is itself described by an energy vs information capacity curve of the sort shown in Figure~\ref{graphs}.
Landauer's principle then reflects a clear application of Temperature as Joules per Bit:
To accommodate one additional bit of information, the environment's capacity must be expanded via a precise quantity of energy.
By definition, this quantity is the temperature $T$ of the environment times one bit.
No unnecessary~$k_B$. No unnecessary~$\ln 2$.

Relatedly, Schumacher advocated~\cite{schumacher2011quantuminfo} that environmental temperature and erasure cost are interchangeable. Namely, one can define temperature \emph{as} the erasure cost.

\subsection{Implications of Redefining the Unit of Temperature}\label{sec:implications}

Adopting the primacy of the bit over the kelvin has implications.
Taken seriously, our proposal leads to the elimination of one of the seven defining constants of the SI, namely, $k_B$.

Entropy can be measured in bits, it can be measured in nats, or, as it has been done historically, it can be measured in $\J/\K$.
A system with an entropy of $1~\NAT$ has $\ln \Omega = 1$ and therefore that same system also has entropy~$k_B$.
Thus
$$
1\ \NAT = k_B = 1.380649 \times 10^{-23}\, \frac{\J}{\K}\,,
$$
and the above expressions relate to the bit via $1~\bit = \ln 2~\NAT$. Alternatively, the constant $k_B$ also plays the role of a conversion factor between the Kelvin scale and the $\J/\bit$ scale:
$$
1 \, \K 
=  \ln 2 \times 1.380649 \times 10^{-23} \, \frac{\J}{\bit} 
\approx 9.5699296 \, \frac{\J}{\Ybit} 
\,.
$$

Seen in this light, Boltzmann's constant is not a fundamental constant, but a conversion factor between those different standards.\footnote{It is often said that the Boltzmann's constant links the average kinetic energy with the temperature, but this contingency in the equipartition theorem is explained by the exponential relation between $S$ and $E$ that arises when we consider degrees of freedom that are quadratic in the energy.} As pointed out by Callen~\cite{callen1991thermodynamics}: ``The constant prefactor [in Boltzmann's entropy] merely determines the scale of [Boltzmann's entropy]; it is chosen to obtain agreement with the Kelvin scale of temperature.''
Therefore, Boltzmann's constant embodies the same parochial dependencies as those of the kelvin.

\begin{figure}[h!]
\centering
\begin{subfigure}[t]{0.38\textwidth} 
\centering
\begin{adjustbox}{width=\textwidth}
\includegraphics[width=0.38\textwidth]{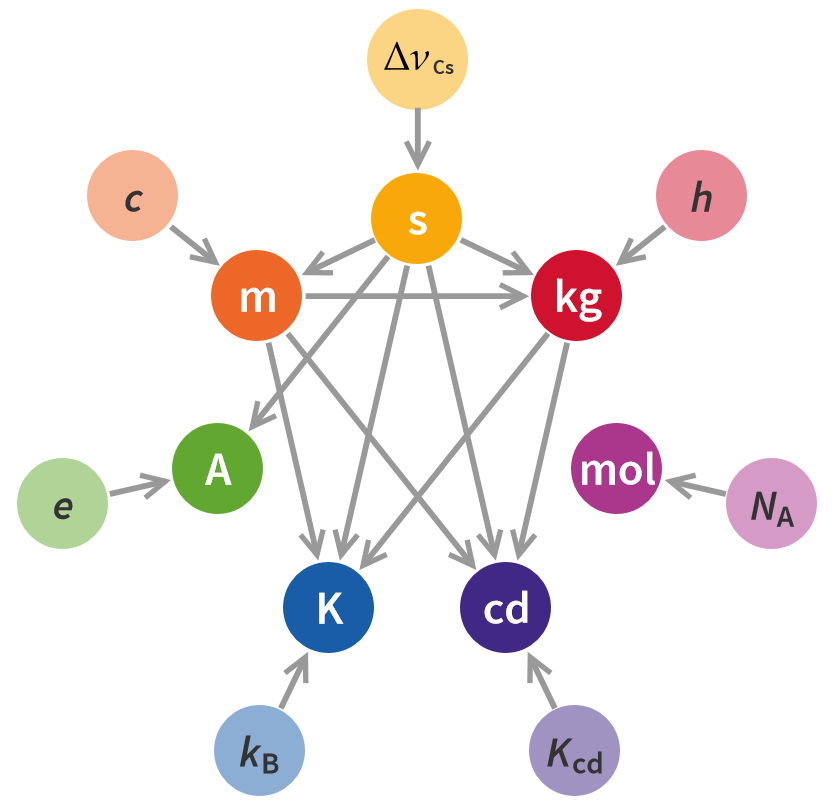}
\end{adjustbox}
\subcaption{The current defining constants\\
		 \hspace{12pt} and SI units, and their\\
		 \hspace{12pt} relationships to one another.}
\label{SI-2019}
\end{subfigure}
\hspace{0.06\textwidth}
\begin{subfigure}[t]{0.38\textwidth} 
\centering
\begin{adjustbox}{width=\textwidth}
\includegraphics[width=0.38\textwidth]{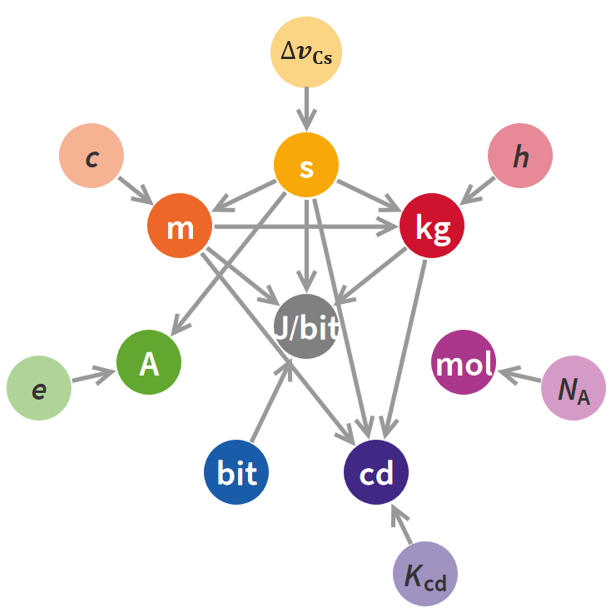}
\end{adjustbox}
\subcaption{Our modification.}
\label{altered-SI}
\end{subfigure}
\captionsetup{width=0.9\linewidth}  
\caption{The result of altering the International System of Units (SI) to recognize the primacy of the $\bit$ and, with it, temperature as $\J/\bit$.
While Boltzmann's constant $k_B$ has been removed, and the Kelvin replaced, no other SI constant or unit is affected by the modification. Note that the units of temperature are no longer a SI base unit. Those of entropy are.}
\label{SI}
\end{figure}

Should the SI adopt the bit instead of the kelvin, the other base units would be left unchanged, as can be readily seen by inspecting Fig.~\ref{SI}. Indeed, no other SI unit depends on the current definition of the kelvin. 
Moreover, the connection between entropy, energy, and temperature is made obvious in the diagram of Fig.~\ref{altered-SI}: Temperature makes the bridge between energy (in $\J$, defined by the arrows coming from the meter, the second, and the kilogram) and entropy (in $\bits$).

Obviously, we do not expect the $\J/\bit$ to feature in everyday discussions about temperature. After all, even if the kelvin is the scientific unit, the common usage of the Fahrenheit and the Celsius scales is still widespread. 
Still, we could describe our preferred pool temperature ($80~^{\circ}\text{F}$) as $2\,869~\J/\Ybit$. As another example, let us consider the entropy of one mole of helium at room temperature and atmospheric pressure. As computed from the Sackur--Tetrode equation~\cite[Chapter~2]{schroeder2000introduction}, its value is 126~$\J/\K$. This corresponds to approximately $13.17~\Ybit$.

\section{Conclusion}

By taking seriously the well-accepted logical priority of entropy over temperature in statistical mechanics, we suggest that entropy ought to find a unit of its own, the bit.
This offers an interdisciplinary interpretation of thermodynamics
centered on the duality between energy and information.
Temperature arises as the nexus between them, linking joules to $\bits$ --- and hence measured in $\J/\bit$.
When a system's entropy is maximal, its information capacity is a useful proxy for its entropy, as they coincide.
The system's temperature is then given by the energy cost to increase the information capacity by one bit.
When the system's entropy is not maximal, it has available energy, or available capacity, like batteries and tapes do.
And viewing the environment as a storehouse of information explains Landauer's ``displacement''; namely, since a bit is never erased, the environment's capacity needs to be increased by one bit to accommodate it, which is achieved at the energy cost (in joules) of $T \times 1 \,\bit$, where $T$ is the temperature of the environment in joules per bit. 

The ideas in this paper were developed under the assumption of classical 
physical laws.
However, this limitation does not detract from the significance of our proposal. 
We conjecture that the interpretation of temperature as energy per unit of information remains valid, regardless of whether the laws of physics are quantum or of any other form yet to be discovered.

Like energy, which takes very different forms across the many domains of physics, information might also be characterized in different ways. 
Investigating them and interlinking information-theoretic notions with many of the traditional concepts of thermodynamics shall be fruitful.

\subsection*{Acknowledgment}
The authors are grateful to Charles H. Bennett, Gilles Brassard, David Deutsch, Paul Erker, Hlér Kristjánsson, Richard MacKenzie, Bryan W. Roberts, Tommaso Toffoli and Maria Violaris for fruitful discussions and comments on earlier versions of this work. The authors also thank the Jude the Obscure Pub for its enlightening atmosphere.
Additionally, the authors acknowledge Gilles Brassard for generously enabling the publication of this work in open access.

Figure~3 was generated from an adaptation of the Mathematica notebook of Emilio Pisanty (dated 2016--2018); made available under the CC BY-SA 4.0 license.

XCR thanks the Fonds National Suisse (FNS) for financial support through the Postdoc.Mobility fellowship program. XCR acknowledges funding from the BMW endowment fund.
CAB's work is supported by the Fonds de recherche du Québec – Nature et technologie, the
FNS, and the Hasler Foundation.
SW acknowledges support from FNS through project No. 214808.

\bibliographystyle{unsrt}
\bibliography{refs.bib}

\begin{thebibliography}{10}

\bibitem{schroeder2000introduction}
Daniel~V Schroeder.
\newblock {\em An Introduction to Thermal Physics}.
\newblock Addison Wesley, 2000.

\bibitem{leibniz1695specimen}
Gottfried~W Leibniz.
\newblock Specimen dynamicum pro admirandis {N}aturae legibus circa corporum
  vires et mutuas actiones detegendis, et ad suas causas revocandis.
\newblock {\em Acta Eruditorum}, 4:145--157, 1695.

\bibitem{young1807course}
Thomas Young.
\newblock {\em A course of lectures on natural philosophy and the mechanical
  arts: in two volumes}, volume~2.
\newblock Johnson, 1807.

\bibitem{carnot1824reflexions}
Sadi Carnot.
\newblock R{\'e}flexions sur la puissance motrice du feu et sur les machines
  propres {\`a} d{\'e}velopper cette puissance.
\newblock {\em Paris: Bachelier}, 1824.

\bibitem{clausius1850ueber}
Rudolf Clausius.
\newblock {\"U}ber die bewegende {K}raft der {W}{\"a}rme und die {G}esetze,
  welche sich daraus f{\"u}r die {W}{\"a}rmelehre selbst ableiten lassen.
\newblock {\em Annalen der Physik}, 155(3):368--397, 1850.

\bibitem{anirban2021constant}
Ankita Anirban.
\newblock A constant by any other name.
\newblock {\em Nature Physics}, 17(5):660--660, 2021.

\bibitem{maruyama2009colloquium}
Koji Maruyama, Franco Nori, and Vlatko Vedral.
\newblock Colloquium: The physics of {M}axwell's demon and information.
\newblock {\em Reviews of Modern Physics}, 81(1):1--23, 2009.

\bibitem{parrondo2015thermodynamics}
Juan~MR Parrondo, Jordan~M Horowitz, and Takahiro Sagawa.
\newblock Thermodynamics of information.
\newblock {\em Nature physics}, 11(2):131--139, 2015.

\bibitem{maxwell1871theory}
James~C Maxwell.
\newblock {\em Theory of heat}.
\newblock Cambridge University Press, 1871.

\bibitem{leff1990maxwell}
Harvey~S Leff and Andrew~F Rex.
\newblock {\em Maxwell's demon: entropy, information, computing}.
\newblock Princeton University Press, 1990.

\bibitem{leff2002maxwell}
Harvey Leff and Andrew~F Rex.
\newblock {\em Maxwell's Demon 2 Entropy, Classical and Quantum Information,
  Computing}.
\newblock CRC Press, 2002.

\bibitem{szilard1929entropieverminderung}
Leo Szilard.
\newblock {\"U}ber die {E}ntropieverminderung in einem thermodynamischen
  {S}ystem bei {E}ingriffen intelligenter {W}esen.
\newblock {\em Zeitschrift f{\"u}r Physik}, 53(11-12):840--856, 1929.

\bibitem{von1955mathematical}
John Von~Neumann.
\newblock {\em Mathematical foundations of quantum mechanics}.
\newblock Princeton University Press, 1955.

\bibitem{gabor1961iv}
Denis Gabor.
\newblock I{V} {L}ight and information.
\newblock In {\em Progress in optics}, volume~1, pages 109--153. Elsevier,
  1961.

\bibitem{brillouin1962science}
Leon Brillouin.
\newblock {\em Science and information theory}.
\newblock Academic Press, 2\textsuperscript{nd} ed., 1962.

\bibitem{landauer1961irreversibility}
Rolf Landauer.
\newblock Irreversibility and heat generation in the computing process.
\newblock {\em IBM journal of Research and Development}, 5(3):183--191, 1961.

\bibitem{bennett1973logical}
Charles~H Bennett.
\newblock Logical reversibility of computation.
\newblock {\em IBM journal of Research and Development}, 17(6):525--532, 1973.

\bibitem{fredkin1982conservative}
Edward Fredkin and Tommaso Toffoli.
\newblock Conservative logic.
\newblock {\em International Journal of Theoretical Physics}, 21(3-4):219--253,
  1982.

\bibitem{bennett1982thermodynamics}
Charles~H Bennett.
\newblock The thermodynamics of computation---a review.
\newblock {\em International Journal of Theoretical Physics}, 21(12):905--940,
  1982.

\bibitem{schumacher2011quantuminfo}
Benjamin Schumacher.
\newblock \emph{In} quantum information and foundations of thermodynamics,
  \emph{ETH Z{\"u}rich, 9-12 August, 2011,
  \url{https://www.video.ethz.ch/conferences/2011/qiftw11/f9bb6f3d-7461-4971-8213-2e3ceb48c0ba.html}
  (accessed 28.02.25)}.

\bibitem{shannon}
Claude~E Shannon.
\newblock A mathematical theory of communication.
\newblock {\em The Bell System Technical Journal}, 27:379--423, 1948.

\bibitem{jaynes1957information}
Edwin~T Jaynes.
\newblock Information theory and statistical mechanics.
\newblock {\em Physical Review}, 106(4):620--630, 1957.

\bibitem{jaynes1957information2}
Edwin~T Jaynes.
\newblock Information theory and statistical mechanics. {II}.
\newblock {\em Physical Review}, 108(2):171--190, 1957.

\bibitem{hanel2014multiplicity}
Rudolf Hanel, Stefan Thurner, and Murray Gell-Mann.
\newblock How multiplicity determines entropy and the derivation of the maximum
  entropy principle for complex systems.
\newblock {\em Proceedings of the National Academy of Sciences},
  111(19):6905--6910, 2014.

\bibitem{shimony1985status}
Abner Shimony.
\newblock The status of the principle of maximum entropy.
\newblock {\em Synthese}, 63(1):35--53, 1985.

\bibitem{toffoli2016entropy}
Tommaso Toffoli.
\newblock Entropy? {H}onest!
\newblock {\em Entropy}, 18(7):247, 2016.

\bibitem{solomonoff1964formal}
Ray~J Solomonoff.
\newblock A formal theory of inductive inference. {Part I}.
\newblock {\em Information and control}, 7(1):1--22, 1964.

\bibitem{Kolmogorov1965}
Andre{\"\i}~N Kolmogorov.
\newblock {Three approaches to the quantitative definition of information}.
\newblock {\em Problemy Peredachi Informatsii}, 1(1):3--11, 1965.

\bibitem{chaitin1966length}
Gregory~J Chaitin.
\newblock On the length of programs for computing finite binary sequences.
\newblock {\em Journal of the ACM}, 13(4):547--569, 1966.

\bibitem{zurek1989algorithmic}
Wojciech~H Zurek.
\newblock Algorithmic randomness and physical entropy.
\newblock {\em Physical Review A}, 40(8):4731, 1989.

\bibitem{gell2010effective}
Murray Gell-Mann and Seth Lloyd.
\newblock Effective complexity.
\newblock In {\em Murray Gell-Mann: Selected Papers}, pages 391--402. World
  Scientific, 2010.

\bibitem{CCC}
{\"{A}}min Baumeler and Stefan Wolf.
\newblock {Causality--Complexity--Consistency: Can space-time be based on logic
  and computation?}
\newblock In {\em Time in Physics}, pages 69--101. Springer, 2017.

\bibitem{holevo1973some}
Alexander~S Holevo.
\newblock Some estimates for the amount of information transmittable by a
  quantum communication channel (in {R}ussian).
\newblock {\em Problemy Predachi Informacii}, 9:3--11, 1973.

\bibitem{brassard1984quantum}
Charles~H Bennett and Gilles Brassard.
\newblock Quantum cryptography: Public key distribution and coin tossing.
\newblock In {\em International conference on computers, systems and signal
  processing}, pages 175--179, 1984.

\bibitem{deutsch1985quantum}
David Deutsch.
\newblock Quantum theory, the {C}hurch--{T}uring principle and the universal
  quantum computer.
\newblock {\em Proceedings of the Royal Society A. Mathematical, Physical and
  Engineering Sciences}, 400(1818):97--117, 1985.

\bibitem{vedral2006introduction}
Vlatko Vedral.
\newblock {\em Introduction to quantum information science}.
\newblock Oxford University Press, USA, 2006.

\bibitem{lubkin1987keeping}
Elihu Lubkin.
\newblock Keeping the entropy of measurement: Szilard revisited.
\newblock {\em International Journal of Theoretical Physics}, 26:523--535,
  1987.

\bibitem{plenio2001physics}
Martin~B Plenio and Vincenzo Vitelli.
\newblock The physics of forgetting: Landauer's erasure principle and
  information theory.
\newblock {\em Contemporary Physics}, 42(1):25--60, 2001.

\bibitem{zurek2003quantum}
Wojciech~H Zurek.
\newblock Quantum discord and {M}axwell's demons.
\newblock {\em Physical Review A}, 67(1):012320, 2003.

\bibitem{rio2011thermodynamic}
L{\'\i}dia~del Rio, Johan {\AA}berg, Renato Renner, Oscar Dahlsten, and Vlatko
  Vedral.
\newblock The thermodynamic meaning of negative entropy.
\newblock {\em Nature}, 474(7349):61--63, 2011.

\bibitem{faist2015minimal}
Philippe Faist, Fr{\'e}d{\'e}ric Dupuis, Jonathan Oppenheim, and Renato Renner.
\newblock The minimal work cost of information processing.
\newblock {\em Nature communications}, 6(1):7669, 2015.

\bibitem{goold2016role}
John Goold, Marcus Huber, Arnau Riera, L{\'\i}dia Del~Rio, and Paul Skrzypczyk.
\newblock The role of quantum information in thermodynamics---a topical review.
\newblock {\em Journal of Physics A: Mathematical and Theoretical},
  49(14):143001, 2016.

\bibitem{lieb1999physics}
Elliott~H Lieb and Jakob Yngvason.
\newblock The physics and mathematics of the second law of thermodynamics.
\newblock {\em Physics Reports}, 310(1):1--96, 1999.

\bibitem{marletto2016constructor}
Chiara Marletto.
\newblock Constructor theory of thermodynamics.
\newblock {\em arXiv preprint arXiv:1608.02625}, 2016.

\bibitem{hulse2018axiomatic}
Austin Hulse, Benjamin Schumacher, and Michael~D Westmoreland.
\newblock Axiomatic information thermodynamics.
\newblock {\em Entropy}, 20(4):237, 2018.

\bibitem{baumeler2022thermodynamics}
{\"A}min Baumeler, Carla Rieger, and Stefan Wolf.
\newblock Thermodynamics as combinatorics: A toy theory.
\newblock In {\em 2022 IEEE Information Theory Workshop (ITW)}, pages 362--367.
  IEEE, 2022.

\bibitem{gibbs1957method}
Josiah~Willard Gibbs.
\newblock A method of geometrical representation of the thermodynamic
  properties by means of surfaces.
\newblock {\em The Collected Works of J. Willard Gibbs, Ph. D., LL. D}, pages
  33--54, 1957.

\bibitem{schrodinger1944life}
Erwin Schr{\"o}dinger.
\newblock {\em What is life?: The physical aspect of the living cell}.
\newblock Cambridge University Press, 1944.

\bibitem{brillouin1953negentropy}
Leon Brillouin.
\newblock The negentropy principle of information.
\newblock {\em Journal of Applied Physics}, 24(9):1152--1163, 1953.

\bibitem{wolf2018second}
Stefan Wolf.
\newblock Second thoughts on the second law.
\newblock In {\em Adventures Between Lower Bounds and Higher Altitudes}, pages
  463--476. Springer, 2018.

\bibitem{baumeler2019free}
{\"A}min Baumeler and Stefan Wolf.
\newblock Free energy of a general computation.
\newblock {\em Physical Review E}, 100(5):052115, 2019.

\bibitem{callen1991thermodynamics}
Herbert~B Callen.
\newblock {\em Thermodynamics and an Introduction to Thermostatistics}.
\newblock John {W}iley \& {S}ons, 1991.

\end{thebibliography}
\end{document}